\documentclass[preprint,nopacs,preprintnumbers,amsmath,amssymb]{revtex4}
\usepackage{graphicx}
\usepackage{amssymb}
\usepackage{xcolor}
\usepackage{fancyhdr}
\usepackage{amsmath}
\usepackage{mathtools}
\usepackage{mathrsfs}
\usepackage{bm}
\usepackage{dsfont}
\usepackage{amsfonts}
\usepackage{bbold}
\begin{document}

\title{Superfluidity of dipolar magnetoexcitons in doped double-layered $\alpha$-${\cal T}_3$ lattice in a strong   magnetic field}
\author{Yonatan Abranyos$^1$, Oleg L. Berman$^2$,  and  Godfrey Gumbs$^1$  }
\affiliation{ $^{1}$Department of Physics and Astronomy, Hunter
College of the City University of New York, 695 Park Avenue, New
York, NY 10065, USA}
 \affiliation{
$^{2}$Physics Department, New York City College of Technology of the
City University of New York, 300 Jay Street, Brooklyn, NY 11201,
USA}
\begin{abstract}
We predict the occurrence of Bose-Einstein condensation and
superfluidity of dipolar magnetoexcitons for a pair of
quasi-two-dimensional spatially separated
$\alpha$-${\cal T}_3$ layers.  We have
solved a two-body problem for an electron and a hole for the model
Hamiltonian for the $\alpha$-${\cal T}_3$ double layer in a magnetic
field. The energy dispersion of collective excitations, the
spectrum of sound velocity, and the effective magnetic mass of
magnetoexcitons are obtained in
 the integer quantum Hall regime for  high magnetic fields. The superfluid density and the temperature of the Kosterlitz-Thouless phase transition
  are probed as functions of the excitonic density, magnetic field and the
inter-layer separation.
\end{abstract}

\date{\today}

\maketitle

\section{Introduction}\label{introduction}
\label{sec1}

The many-particle systems of dipolar (indirect) excitons, formed by spatially separated electrons and holes, in semiconductor coupled quantum wells (CQWs)  in a magnetic field $B$, as well as in the absence of magnetic field,  have attracted considerable attention.  This interest has been  generated in large part by the possibility of Bose-Einstein condensation (BEC) and superfluidity of dipolar excitons, which can be observed  as persistent electrical currents in each quantum  well, and also through coherent optical properties~\cite{Lozovik,Snoke,Butov,Eisenstein}. Recent  progress in theoretical and experimental studies of the superfluidity of  dipolar excitons in CQWs was reviewed in Ref.~\cite{Snoke_review}.

\medskip
\par

\medskip
\par
Recently,  a number of experimental and theoretical studies were dedicated to graphene, and the condensation of electron-hole pairs, formed by spatially separated electrons and holes,  in a pair of parallel graphene layers.   These investigations were reported in Refs.~\cite{BLG,Sokolik,Bist,BKZg,Perali}. Since the exciton binding energies in novel two-dimensional (2D) semiconductors is quite large, both BEC and superfluidity of dipolar excitons in double layers of transition-metal dichalcogenides (TMDCs)~\cite{Fogler,MacDonald_TMDC,BK,BK2} and phosphorene~\cite{BGK,Peeters} have been discussed. In high  magnetic fields, 2D excitons  referred to as {\it magnetoexcitons} exist in a much wider temperature range, since the magnetoexciton binding energies increase as the magnetic field is increased ~\cite{Lerner,Paquet,Kallin,Yoshioka,Ruvinsky,Ulloa,Moskalenko}.

\medskip
\par

Lately, there has been growing interest in the electronic properties of the $\alpha$-${\cal T}_3$ lattice for its surprising fundamental physical phenomena as well as its promising applications in solid state devices\cite{f1,f2,f3,f4,f5,f6,f7,f8,f9,RKKY,t1,t2,t3}. For a review of artificial flat band systems, see Ref.\  \cite{Review}. Raoux, et al. \cite{f1}  proposed  that an $\alpha$-${\cal T}_3$ lattice could be assembled  from cold fermionic atoms  confined to an optical lattice by means of three pairs of laser beams for the optical dice lattice ($\alpha=1$) \cite{21}. A model of  this structure, consists of an AB-honeycomb lattice (the rim)  like that in graphene which is combined with C atoms at the center/hub of each hexagon.  A parameter $\alpha$ is then introduced to represent the ratio of the hopping integral between the hub and the rim to that around the rim of the hexagonal lattice.  By dephasing one of the three pairs of laser beams, one  could possibly vary the  parameter $0<\alpha =\tan\ \phi<1$.

\medskip
\par

We consider a pair of parallel $\alpha$-${\cal T}_3$
layers separated by an insulating slab (e.g., SiO$_2$ or hexagonal
boron nitride ($h$-BN)) in a strong perpendicular magnetic field.
The equilibrium system of local pairs of electrons and holes,
spatially separated on these parallel $\alpha$-${\cal
T}_3$ layers, correspondingly, can be created by varying the
chemical potential using a bias voltage between the two
$\alpha$-${\cal T}_3$ layers or between two gates
located near the respective $\alpha$-${\cal T}_3$ 2D
sheets  (for simplicity, we also call these equilibrium local
electron-hole (e-h) pairs as dipolar magnetoexcitons). 
In case 1 described above, a dipolar magnetoexciton is formed by an
electron in the Landau level $1$ and a hole in the Landau level
$-1$. Dipolar magnetoexcitons with spatially separated electrons and
holes can also be created by laser pumping  and by applying a
perpendicular electric field  as it is done for
CQWs~\cite{Snoke,Butov,Eisenstein}. In case 2 described
above, a dipolar magnetoexciton is formed by an electron in the
Landau level $1$ and hole in the Landau level $0$. We assume that
the system is in a quasi-equilibrium state. We investigate the
collective properties and propose the occurrence of superfluidity of
dipolar excitons in $\alpha$-${\cal T}_3$ double
layers in high magnetic field for both cases 1 and 2. 
We assume that the dilute system of magnetoexcitons forms a
weakly-interacting Bose gas.

\medskip
\par

Our decision to investigate dipolar magnetoexcitons in a double layer versus direct magnetoexcitons in a monolayer
was driven by the fact that the e-h recombination due to tunneling of electrons and holes between monolayers in a double
layer is suppressed by the dielectric barrier, which is placed between two monolayers~\cite{BLG}.
Therefore, the dipolar magnetoexcitons, formed by electrons and holes, located in two separate  $\alpha$-${\cal T}_3$ layers,
have a longer lifetime than the direct magnetoexcitons in a single $\alpha$-${\cal T}_3$ layer. Moreover, due to the interlayer separation $D$,
 dipolar magnetoexcitons both in the ground and excited states have non-zero electrical dipole moments.
 The dipole moments of the dipolar magnetoexciton produce a long-range dipole-dipole repulsion between magnetoexcitons,
 which leads to larger sound velocity and, consequently  higher critical temperature for superfluidity of the dipolar magnetoexcitons in a double layer
 compared with the direct excitons in a monolayer having the same magnetoexciton densities.

\medskip
\par
The rest of the paper is organized in the following way. In Sec. \ \ref{sec2}, the model for electrons
 in an $\alpha$-${\cal T}_3$ monolayer in a perpendicular magnetic field is reviewed so as to establish our notation.
 In Sec. \ref{sec3}, the two-body problem for an electron and a hole, spatially separated in two parallel $\alpha$-${\cal T}_3$ monolayers
  in a perpendicular magnetic field, is formulated, and the corresponding eigenenergies and wave functions are derived. In Sec. \ref{sec4},
   the effective masses and binding energies for isolated dipolar magnetoexcitons in the $\alpha$-${\cal T}_3$ double layer are obtained.
   The collective properties and superfluidity of the weakly interacting Bose gas of dipolar magnetoexcitons in the $\alpha$-${\cal T}_3$
   double layer are investigated in Sec. \ref{sec5}. Our conclusions are presented in Sec.~\ref{sec6}.

\section{$\alpha$-${\rm T}_3$ Model in a magnetic Field}
\label{sec2}

In the absence of magnetic field,  the Hamiltonian near the K  point  is given by

\begin{eqnarray}
H=\hbar v_F
\begin{pmatrix}
0 &(p_x+ip_y)\cos\phi&  0  \\
(p_x-ip_y)\cos\phi & 0 &(p_x+ip_y)\sin\phi \\
0 & (p_x-ip_y)\sin\phi & 0\\
\end{pmatrix}
\end{eqnarray}
with $v_F$ the Fermi velocity and the  parameter  $\alpha=\tan\phi$ describing the  strength of the hopping to the central C-atoms. In the presence of a magnetic field ${\bf B}=B\hat{\bf e}_z$ parallel to  the $z$-axis, we use the Landau gauge ${\bf A}=xB\hat{\bf e}_y$ and with minimal coupling ${\bf p}\to {\bf p}\pm e{\bf A}$ for electrons and holes respectively. In addition, we have Zeeman splitting and a term for pseudo-spin splitting. For now we ignore the Zeeman and pseudo-spin splitting. With the minimal coupling substitution in the Landau gauge we obtain the Hamiltonian for electrons and holes using the annihilation operators for an electron and hole as follows

\begin{eqnarray}
&&c_{\pm}=\frac{1}{\sqrt{2\hbar eB}}\Big(p_x+i(p_y\mp exB\Big)\\
&&[c_+,c^\dagger_+]=1,\;\;\;[c_-,c^\dagger_-]=1,\;\;\;c^\dagger_+c_+=\hat{n}_+,\;\;\;
c^\dagger_-c_-=\hat{n}_-\nonumber
\end{eqnarray}
The Hamiltonian for the $K$ and  $K'$ valleys are given in terms of these operators

\begin{eqnarray}
H^{eK}_{kin}=\sqrt{2}\frac{\hbar v_F}{r_B}
\begin{pmatrix}
0 &c_+\cos\phi&  0  \\
c^\dagger_+\cos\phi & 0 &c_+\sin\phi \\
0 & c^\dagger_+\sin\phi & 0\\
\end{pmatrix},\;\;\;\;
H^{eK'}_{kin}=\sqrt{2}\frac{\hbar v_F}{r_B}
\begin{pmatrix}
0 &c^\dagger_+\cos\phi&  0  \\
c_+\cos\phi & 0 &c^\dagger_+\sin\phi \\
0 & c_+\sin\phi & 0\\
\end{pmatrix}
\end{eqnarray}
The two independent modes, around $K$ and and $K'$ describing the full low energy Hamiltonian takes the form

\begin{eqnarray}
H^e_{kin}=\sqrt{2}\frac{\hbar v_F}{r_B}
\begin{pmatrix}
0 &c_+\cos\phi&  0 &0 &0 &0 \\
c^\dagger_+\cos\phi & 0 &c_+\sin\phi  & 0 &0 &0\\
0 & c^\dagger_+\sin\phi & 0 &0 & 0&0\\
0& 0& 0&0 &c^\dagger_+\cos\phi&  0  \\
0& 0& 0&c_+\cos\phi & 0 &c^\dagger_+\sin\phi \\
0& 0& 0&0 & c_+\sin\phi & 0\\
\end{pmatrix}
\label{hamiltonk}
\end{eqnarray}
and $r_B=\frac{1}{\sqrt{\hbar eB}}$ is a magnetic length scale

\medskip
\par

The energy eigenvalues are obtained in a similar  way to that for  graphene  \cite{Iyengar}, and we obtain for an electron in the $K$ valley

\begin{eqnarray}
\sqrt{2}\frac{\hbar v_F}{r_B}
\begin{pmatrix}
0 &c_+\cos\phi&  0  \\
c^\dagger_+\cos\phi & 0 &c_+\sin\phi \\
0 & c^\dagger_+\sin\phi & 0\\
\end{pmatrix}
\begin{pmatrix}
a^K_n(\phi)|n-2\rangle  \\
\pm|n-1\rangle \\
b^K_n(\phi)|n\rangle \\
\end{pmatrix}
=\varepsilon_{n,s}\begin{pmatrix}
a^K_n(\phi)|n-2\rangle  \\
\pm|n-1\rangle \\
b^K_n(\phi)|n\rangle \\
\end{pmatrix}
  \  .
    \end{eqnarray}
where $|m?$ is a harmonic oscillator wave function. We have a similar equation for the $K^\prime$ valley giving the energy eigenvalues

\begin{eqnarray}
&&\varepsilon_{n,s}=\sqrt{2}\text{sign}(n)\frac{\hbar v_F}{r_B}\sqrt{n-\frac{1}
{2}\left(1+\eta\cos2\phi\right)},
\;\;\;\varepsilon_{n,0}=0\text{   flat band},\;\;\;n=2,\,3,\cdots
\label{energy}
\end{eqnarray}
Here $\eta=\pm1$ with $\eta=1$ for the $K$ valley and $\eta=-1$ for $K'$ valley. The corresponding energy eigenstates for $n=2,\,3,\cdots$ are

\begin{eqnarray}
&&|\psi^K_{\pm,n}\rangle=\frac{1}{\sqrt{2}}\begin{pmatrix}
a^K_n(\phi)|n-2\rangle  \\
\pm|n-1\rangle  \\
b^K_n(\phi)|n\rangle)  \\
\end{pmatrix},\;\;\;\;\;
|\psi^{K'}_{\pm,n}\rangle=\frac{1}{\sqrt{2}}\begin{pmatrix}
a^{K'}_n(\phi)|n\rangle \\
\pm|n-1\rangle \\
b^{K'}_n(\phi)|n-2) \\
\end{pmatrix}  \  .
\end{eqnarray}
In this notation,

\[a^K_n(\phi)=\sqrt{\frac{(n-1)\cos^2\phi}{n-\cos^2\phi}},\;
b^K_n(\phi)=\sqrt{\frac{n\sin^2\phi}{n-\cos^2\phi}},\;
a^{K'}_n(\phi)=-\sqrt{\frac{n\cos^2\phi}{n-\sin^2\phi}},\;
b^{K'}_n(\phi)=\sqrt{\frac{(n-1)\sin^2\phi}{n-\sin^2\phi}}\]
 For the flat-band with   $\varepsilon_{n,0}=0$,  the eigenstates are given by

\begin{eqnarray*}
&&|\psi^K_{0,n}\rangle=\frac{1}{\sqrt{2}}\begin{pmatrix}
b^K_n(\phi)|n-2\rangle  \\
0 \\
a^K_n(\phi)|n\rangle)  \\
\end{pmatrix},\;\;\;\;\;
|\psi^{K'}_{0,n}\rangle=\frac{1}{\sqrt{2}}\begin{pmatrix}
b^{K'}_n(\phi)|n\rangle \\
0\\
a^{K'}_n(\phi)|n-2) \\
\end{pmatrix}
\end{eqnarray*}
We treat the lowest state $n=1$ separately. In this case,  the eigenvalue problem is
\begin{eqnarray}
H^{eK}_{kin}|\Psi\rangle=\frac{\hbar v_F}{r_B}\sqrt{2}
\begin{pmatrix}
0 &c_+\cos\phi&  0  \\
c^\dagger_+\cos\phi & 0 &c_+\sin\phi \\
0 & c^\dagger_+\sin\phi & 0\\
\end{pmatrix}
\begin{pmatrix}
0  \\
\alpha|0\rangle \\
\beta|1\rangle \\
\end{pmatrix}
=\varepsilon\begin{pmatrix}
0  \\
\alpha|0\rangle \\
\beta|1\rangle \\
\end{pmatrix}
\end{eqnarray}
The energy eigenvalues and eigenstates are
\begin{eqnarray}
&&|\psi^{K}_{\pm,1}\rangle=\frac{1}{\sqrt{2}}
\begin{pmatrix}
0  \\
\pm|0\rangle \\
|1\rangle \\
\end{pmatrix},\;\;\;
\varepsilon_{1,\pm}=\pm\sqrt{2}\frac{\hbar v_F}{r_B}\sin\phi\\
&&|\psi^{K'}_{\pm,1}\rangle=\frac{1}{\sqrt{2}}
\begin{pmatrix}
|1\rangle  \\
\pm|0\rangle \\
0 \\
\end{pmatrix},\;\;\;
\varepsilon_{1,\pm}=\pm\sqrt{2}\frac{\hbar v_F}{r_B}\cos\phi,
\end{eqnarray}
There is no flat band wave function associated with the $n=1$.

\section{Two-body problem for an electron and a hole in the $\alpha-T_3$ double layer in a perpendicular magnetic field}
\label{sec3}

We first consider the Hamiltonian for a non-interacting electron-hole pair excluding the Coulomb interaction. We choose the electron-hole state belonging to a single , $K$ valley. In general, the magnetoexciton state is a superposition of $K$ and $K^\prime$ valley states. Therefore, we will confine our states to the subspace of $K$ valley states in Eq.\ref{hamiltonk} (upper right $3\times3$ block), i.e.,

\[H=H^e_{kin}\otimes\mathds{1}_h+\mathds{1}_e\otimes H^h_{kin}\]
In matrix form, we have

\begin{eqnarray}
H^{e-h}_{kin}=H_{kin}=\frac{\hbar v_F}{r_B}\sqrt{2}
\begin{pmatrix}
H^h_{kin} & c_+\cos\phi\mathds{1}_h  &  0  \\
c^\dagger_+\cos\phi\mathds{1}_h & H^h_{kin}  &c_+\sin\phi\mathds{1}_h \\
0 & c^\dagger_+\sin\phi\mathds{1}_h & H^h_{kin}\\
\end{pmatrix} \  .
\end{eqnarray}
This is a $9\times9$ matrix where each entry above is a $3\times3$-matrix.

\begin{figure} [h]
\includegraphics[width=.60\linewidth]{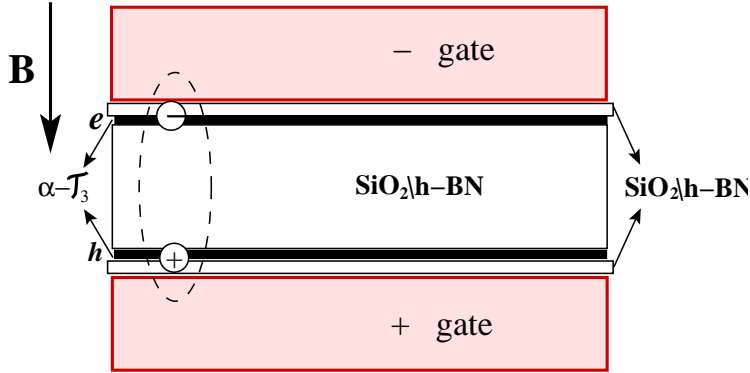}
\caption{Schmematic illustration of a dipolar magnetoexciton on a pair of $\alpha$-${\cal T}_3$ layers embedded in an insulating material. A uniform perpendicular magnetic field ${\bf B}$ is
applied, and negative and positive biases are attached to the layers in the $xy$-plane.}
\label{FIG:1}
\end{figure}
A  schematic illustration of a dipolar magnetoexciton, which is a bound state of  a spatially separated electron and a hole, located on a pair of $\alpha$-${\cal T}_3$ layers embedded in an insulating material in a perpendicular magnetic field ${\bf B}$, is depicted in Fig.~\ref{FIG:1}. In the case  of non-interacting excitons, the eigenvalues are additive and we obtain

\begin{eqnarray}
&&\varepsilon_{n_+,n_-}=\frac{\hbar v_F}{r_B}\sqrt{2}\Bigg(\text{sgn}(n_+)\sqrt{n_+-\frac{1}{2}\left(1+\eta\cos2\phi\right)}\nonumber\\
&&\times \text{sgn}(n_-)\sqrt{n_+-\frac{1}{2}\left(1-\eta\cos2\phi\right)}\Bigg) \   .
\end{eqnarray}
The general eigenstates are a superposition of product states of the form

\begin{eqnarray*}
&&|\Psi_{n_+,n_-}\rangle =
|\psi_{n_+}\rangle\otimes|\psi_{n_-}\rangle\\
&&|\Psi\rangle=\sum_{n_+,n_-}a(n_+,n_-)|\Psi_{n_+,n_-}\rangle  \  .
\end{eqnarray*}

\medskip
\par

We now rewrite the Hamiltonian  in the center-of-mass (CM) and relative coordinates. The energy of indirect excitons is obtained when a substrate is sandwiched between  a double-layer of $\alpha$-${\cal T}_3$. We then have the Coulomb term $u({\bf r}_e-{\bf r}_h)$ between electron at ${\bf r}_e$ and hole at ${\bf r}_h$. The magneto-exciton Hamiltonian is given by

\begin{eqnarray}
H&&=H^e_{kin}\otimes\mathds{1}_h+\mathds{1}_e\otimes H^h_{kin}+u({\bf r}_e-{\bf r}_h)\otimes\mathds{1}
\end{eqnarray}
\begin{eqnarray*}
&&=\hbar v_F
\begin{pmatrix}
0 &\cos\phi[p_{ex}+i(p_{ey}+eBx_e)]&  0  \\
\cos\phi[p_{ex}-i(p_{ey}+eBx_e)] & 0  & \sin\phi[p_{ex}+i(p_{ey}+eBx_e)] \\
0 & \sin\phi[p_{ex}-i(p_{ey}+eBx_e)] & 0\\
\end{pmatrix}\otimes\mathds{1}_h\\
&&\\
&&+\hbar v_F\mathds{1}_e\otimes
\begin{pmatrix}
0 &\cos\phi[p_{hx}-i(p_{hy}-eBx_h)]&  0  \\
\cos\phi[p_{hx}+i(p_{hy}-eBx_h)] & 0  & \sin\phi[p_{hx}-i(p_{hy}-eBx_h)] \\
0 & \sin\phi[p_{hx}+i(p_{hy}-eBx_h)] & 0\\
\end{pmatrix}
\end{eqnarray*}
We go to  center of mass and relative coordinate system as follows.
\[{\bf p}_{e/h}={\bf P}/2\pm{\bf p},\;\;\;\;{\bf r}_{e/h}={\bf R}\pm{\bf r}/2\]

We define the pseudospin-1 operators $S_x(\phi)$ and $S_y(\phi)$ as

\begin{eqnarray*}
S_{x}(\phi)=
\begin{pmatrix}
0 &\cos\phi&  0  \\
\cos\phi & 0  & \sin\phi \\
0 & \sin\phi & 0\\
\end{pmatrix}\;\;\;\;\;\;
S_{y}(\phi)=i
\begin{pmatrix}
0 &-\cos\phi&  0  \\
\cos\phi & 0  & -\sin\phi \\
0 & \sin\phi & 0\\
\end{pmatrix}
\end{eqnarray*}
The total Hamiltonian is then given by

\begin{eqnarray*}
H=&&\hbar v_F\Big(P_xm^+_x(\phi)+2p_xm^-_x(\phi)+(P_y-eBx)m^-_y(\phi)+2p_ym^+_y(\phi)\\
&&+2(p_y+eBX)m^+_y(\phi)\Big)+u({\bf r})\otimes\mathds{1}
\end{eqnarray*}
In this notation,

\begin{eqnarray*}
&&m^{\pm}_x(\phi)=\frac{1}{2}\left[S_x(\phi)\otimes\mathds{1}_h
\pm\mathds{1}_e\otimes S_x(\phi)\right]\\
&&m^{\pm}_y(\phi)=\frac{1}{2}\left[S_y(\phi)\otimes\mathds{1}_h
\pm\mathds{1}_e\otimes S_y(\phi)\right] \  .
\end{eqnarray*}
The unitary operator \(U=e^{ieBXy}\)  transforms the Hamiltonian

\begin{eqnarray*}
&&U^\dagger H U=\hbar v_F\Big(2{\bf p}\cdot{\bf m}^{-}(\phi)+(\bf{P}-\hat{z}\times{\bf r})
\cdot{\bf m}^{+}(\phi)\Big)+u({\bf r})\otimes\mathds{1}
\end{eqnarray*}
shifting ${\bf r}\to {\bf r-\hat{z}}\times{\bf P}$, moves the ${\bf P}$ dependence to the potential energy.
 This gives the same form of two-particle wave function as in \cite{BLG}.
  Here are the states and the energies in the notation and form as \cite{BLG}  (see Eqs. 4 and  6).
   The nine-component wave function is written as follows. Note  the symbols have the same definitions as  \cite{BLG}. We have
\begin{equation}
\Psi({\bf R}, {\bf r})=\exp\left[i\left({\bf P}+\frac{e}{2c}[{\bf B}\times{\bf r}]\right)
\cdot\frac{{\bf R}}{\hbar}\right]\tilde\Phi({\bf r}-{\bf {\bm\rho}}_0) \  .
\end{equation}

\subsection{Magnetoexciton states for $n_+,\,n_-=2,\,3,\cdots$}

We assume that both electrons and holes  are in the $K$ valley and  we first consider  $n_+=2,\,3,\cdots$ and $n_-=2,\,3,\cdots$ states. Therefore, for a magnetoexciton state, we use the tensor product of the above states for the electron-hole wave function. We express the wave function using

\begin{eqnarray}
&&\Phi_{n_+,n_-}({\bf r})=
(2\pi)^{-1/2}2^{-|m|/2}\frac{\tilde{n}!}{\sqrt{n_+!n_-!}}\frac{1}{r_B}\text{sgn}(m)^m
\frac{r^{|m|}}{r^{|m|}_B}\exp\left[-im\phi-\frac{r^2}{4r^2_B} \right]
L^{|m|}_{\tilde{n}}\left(\frac{r^2}{2r^2_B}\right) \   ,
\end{eqnarray}
where $m=\left|n_+-n_-\right|$, $\tilde{n}=\min(n_+,n_-)$ and $L$ are Laguerre polynomials. We have

\begin{eqnarray}
&&\psi_{n_+}({\bf r}_e)\otimes\psi_{n_-}({\bf r}_h)
=\frac{1}{2}\begin{pmatrix}
a_{n_+}(\phi)\Phi_{n_+-2}({\bf r}_e)  \\
\Phi_{n_+-1}({\bf r}_e)  \\
b_{n_+}(\phi)\Phi_{n_+}({\bf r}_e)  \\
\end{pmatrix}
\otimes\begin{pmatrix}
 a_{n_-}(\phi)\Phi_{n_--2}({\bf r}_h) \\
-\Phi_{n_--1}({\bf r}_h) \\
b_{n_-}(\phi)\Phi_{n_-}({\bf r}_h) \\
\end{pmatrix} \  .
\end{eqnarray}
This leads to the following wave function in the CM coordinate frame of reference.

\begin{eqnarray}
&&\tilde\Phi_{n_+,n_-}({\bf r})=\frac{1}{2}\begin{pmatrix}
a_{n_+}(\phi)a_{n_-}(\phi)\Phi_{n_+-2,n_--2}({\bf r})  \\
-a_{n_+}(\phi)\Phi_{n_+-2,n_--1}({\bf r})  \\
a_{n_+}(\phi)b_{n_-}(\phi)\Phi_{n_+-2,n_-}({\bf r})  \\
a_{n_-}(\phi)\Phi_{n_+-1,n_--2}({\bf r})  \\
-\Phi_{n_+-1,n_--1}({\bf r})  \\
b_{n_-}(\phi)\Phi_{n_+-1,n_-}({\bf r})  \\
b_{n_+}(\phi)a_{n_-}(\phi)\Phi_{n_+,n_--2}({\bf r})  \\
-b_{n_+}(\phi)\Phi_{n_+,n_--1}({\bf r})  \\
b_{n_+}(\phi)a_{n_-}(\phi)\Phi_{n_+,n_-}({\bf r})  \\
\end{pmatrix} \  .
\label{mexcitongeneral}
\end{eqnarray}

\subsection{Landau-Levels for $ n_{\pm}=1$}

The $n_{\pm}=1$ Landau-Levels are treated separately as described below. We express the eigenvalue
problem as

\begin{eqnarray}
H^e_{kin}|\Psi\rangle=\gamma_B
\begin{pmatrix}
0 &c_+\cos\phi&  0  \\
c^\dagger_+\cos\phi & 0 &c_+\sin\phi \\
0 & c^\dagger_+\sin\phi & 0\\
\end{pmatrix}
\begin{pmatrix}
0  \\
\alpha|0\rangle \\
\beta|1\rangle \\
\end{pmatrix}
=\varepsilon\begin{pmatrix}
0  \\
\alpha|0\rangle \\
\beta|1\rangle \\
\end{pmatrix}
\end{eqnarray}
and we have a similar equation for a hole from which  the states are  given by

\begin{eqnarray}
&&
\psi_{n_+=1}({\bf r}_e)
=\frac{1}{\sqrt{2}}\begin{pmatrix}
0  \\
\pm\Phi_{0}({\bf r}_e)  \\
\Phi_{1}({\bf r}_e)  \\
\end{pmatrix},
\;\;\;\;\;\psi_{n_-=1}({\bf r}_h)
=\frac{1}{\sqrt{2}}\begin{pmatrix}
0 \\
\pm\Phi_{0}({\bf r}_h)  \\
 \Phi_{1}({\bf r}_h) \\
\end{pmatrix}
\end{eqnarray}
We note that the $n_{\pm}=1$ states are independent of the hopping parameter  $\phi$

\subsection{Magnetoexciton states for $n_+=2,\,3,\cdots\;\;\;n_-=1$}

\begin{eqnarray}
&&\psi_{n_+}({\bf r}_e)\otimes\psi_{n_-}({\bf r}_h)
=\frac{1}{2}\begin{pmatrix}
a_{n_+}(\phi)\Phi_{n_+-2}({\bf r}_e)  \\
\Phi_{n_+-1}({\bf r}_e)  \\
b_{n_+}(\phi)\Phi_{n_+}({\bf r}_e)  \\
\end{pmatrix}
\otimes\begin{pmatrix}
 0 \\
-\Phi_{0}({\bf r}_h) \\
\Phi_{1}({\bf r}_h) \\
\end{pmatrix}
\end{eqnarray}
This leads to the following wave function in the CM coordinate system.
\begin{eqnarray}
&&\tilde\Phi_{n_+,n_-}({\bf r})=\frac{1}{2}\begin{pmatrix}
0  \\
-a_{n_+}(\phi)\Phi_{n_+-2,0}({\bf r})  \\
a_{n_+}(\phi)\Phi_{n_+-2,1}({\bf r})  \\
0  \\
-\Phi_{n_+-1,0}({\bf r})  \\
\Phi_{n_+-1,1}({\bf r})  \\
0  \\
-b_{n_+}(\phi)\Phi_{n_+,0}({\bf r})  \\
b_{n_+}(\phi)\Phi_{n_+,1}({\bf r})  \\
\end{pmatrix}
\label{mexcitonn1}
\end{eqnarray}

\subsection{Magnetoexciton states for $n_+=1,\;\;\;n_-=2,\,3,\cdots$}

\begin{eqnarray}
&&\psi_{n_+}({\bf r}_e)\otimes\psi_{n_-}({\bf r}_h)
=\frac{1}{2}
\begin{pmatrix}
 0 \\
\Phi_{0}({\bf r}_e) \\
\Phi_{1}({\bf r}_e) \\
\end{pmatrix}\otimes
\begin{pmatrix}
a_{n_-}(\phi)\Phi_{n_--2}({\bf r}_h)  \\
-\Phi_{n_--1}({\bf r}_h)  \\
b_{n_-}(\phi)\Phi_{n_-}({\bf r}_h)  \\
\end{pmatrix}
\end{eqnarray}
This leads to the following wave function in the CM coordinate system.
\begin{eqnarray}
&&\tilde\Phi_{1,n_-}({\bf r})=\frac{1}{2}\begin{pmatrix}
0  \\
0  \\
0  \\
a_{n_-}(\phi)\Phi_{0,n_{-}-2}({\bf r}) \\
-\Phi_{0,n_{-}-1}({\bf r})  \\
b_{n_-}(\phi)\Phi_{0,n_{-}}({\bf r})  \\
a_{n_-}(\phi)\Phi_{1,n_{-}-2}({\bf r}) \\
-\Phi_{1,n_{-}-1}({\bf r})  \\
b_{n_-}(\phi)\Phi_{1,n_{-}}({\bf r})  \\
\end{pmatrix}
\label{mexcitonn101}
\end{eqnarray}
\subsection{Magnetoexciton states for $n_+=1\;\;n_-=1$}
We note that  for $n=0$ we have no valence or conduction band states, but there is a flat band described by

\begin{eqnarray}
&&\psi_{n_+=1}({\bf r}_e)\otimes\psi_{n_-=1}({\bf r}_h)
=\frac{1}{2}
\begin{pmatrix}
 0 \\
+\Phi_{0}({\bf r}_e) \\
\Phi_{1}({\bf r}_e) \\
\end{pmatrix}
\otimes
\begin{pmatrix}
 0 \\
-\Phi_{0}({\bf r}_h) \\
\Phi_{1}({\bf r}_h) \\
\end{pmatrix}
\end{eqnarray}
which yields the following wave function in the CM frame of reference as

\begin{eqnarray}
&&\tilde\Phi_{n_+=1,n_-=1}({\bf r})=\frac{1}{2}\begin{pmatrix}
0  \\
0  \\
0  \\
0  \\
-\Phi_{0,0}({\bf r})  \\
\Phi_{0,1}({\bf r})  \\
0  \\
\Phi_{1,0}({\bf r})  \\
\Phi_{1,1}({\bf r})  \\
\end{pmatrix} \  .
\label{mexciton11}
\end{eqnarray}

\subsection{Magnetoexciton states for $n_+=1\;\;n_-=0$}

For the case when $n=0$, there is neither a valence nor conduction band, but there is a flat-band. We now consider an electron in $n_+=1$ and a hole in the flat band with $n_-=0$.   The $n=0$ state

\begin{eqnarray}
&&\psi_{n_-=0}({\bf r}_h)\
=
\begin{pmatrix}
 0 \\
0 \\
\Phi_{0}({\bf r}_h) \\
\end{pmatrix} \  .
\end{eqnarray}
The corresponding exciton state becomes

\begin{eqnarray}
&&\psi_{n_+=1}({\bf r}_e)\otimes\psi_{n_-=0}({\bf r}_h)
=\frac{1}{\sqrt{2}}
\begin{pmatrix}
 0 \\
\Phi_{0}({\bf r}_e) \\
\Phi_{1}({\bf r}_e) \\
\end{pmatrix}
\otimes
\begin{pmatrix}
 0 \\
0 \\
\Phi_{0}({\bf r}_h) \\
\end{pmatrix} \   .
\end{eqnarray}
This leads to the following wave function in the CM coordinate system.

\begin{eqnarray}
&&\tilde\Phi_{n_+=1,n_-=0}({\bf r})=\frac{1}{\sqrt{2}}\begin{pmatrix}
0  \\
0  \\
0  \\
0  \\
0  \\
\Phi_{0,0}({\bf r})  \\
0  \\
0  \\
\Phi_{1,0}({\bf r})  \\
\end{pmatrix} \  .
\label{mexciton1101}
\end{eqnarray}

\section{Isolated dipolar magnetoexciton}
\label{sec4}

For case 1, we calculate the magnetoexciton energy using the
expectation value for an electron in Landau level $1$ and a hole in
level $1$. In high magnetic field, the magnetoexciton is constructed
from an electron and hole in the lowest Landau level with the following nine-component wave function having relative
coordinates

\begin{eqnarray}\label{waveex1}
\tilde{\Phi}_{1,1}(\mathbf{r}) =    \left(
\begin{array}{c}
0\\
0\\
0\\
0\\
- \Phi_{0,0}(\mathbf{r})\\
\Phi_{0,1}(\mathbf{r})\\
0\\
 \Phi_{1,0}(\mathbf{r})\\
\Phi_{1,1}(\mathbf{r})
\end{array}\right)\ .
\end{eqnarray}

\medskip
\par
For case 2, we calculate the magnetoexciton energy using the expectation value for an electron in Landau level $1$ and a hole in level $0$. We have

\begin{eqnarray}\label{waveex2}
\tilde{\Phi}_{1,0}(\mathbf{r}) =    \left(
\begin{array}{c}
0\\
0\\
0\\
0\\
0\\
\Phi_{0,0}(\mathbf{r})\\
0\\
0\\
\Phi_{1,0}(\mathbf{r})\
\end{array}\right)\ .
\end{eqnarray}
The 2D harmonic oscillator  eigenfunctions $\Phi_{n_{e},n_{h}}(\mathbf{r})$ are given by  \cite{Iyengar}

\begin{eqnarray}\label{electron_S}
\Phi_{n_{e},n_{h}}(\mathbf{r}) =
(2\pi)^{-1/2}2^{-|m|/2}\frac{\tilde{n}!}{\sqrt{n_{1}!n_{2}!}}
\frac{1}{r_{B}} \mathrm{sgn}(m)^{m}\frac{r^{|m|}}{r_{B}^{|m|}}
\exp\left[-im\phi - \frac{r^{2}}{4r_{B}^{2}}\right]
L_{\tilde{n}}^{|m|}\left(\frac{r^{2}}{2r_{B}^{2}}\right) \ ,
\end{eqnarray}
where $r_{B} = \sqrt{\hbar/(eB)}$ is the magnetic length, $L_{\tilde{n}}^{|m|}(x)$ denotes Laguerre polynomials; $m = n_{e} -n_{h}$; $\tilde{n} = \min(n_{e},n_{h})$, and $\mathrm{sgn}(m)^{m} = 1$ for $m=0$.

\medskip
\par
The magnetoexciton energy in high magnetic field can be calculated
by employing perturbation theory with respect to Coulombic
electron-hole attraction analogously to 2D quantum wells with finite
electron and hole masses~\cite{Lerner}. This approach
allows us to derive the spectrum of isolated dipolar magnetoexcitons
with spatially separated electrons and holes in the
$\alpha$-${\cal T}_3$ double layer. For the
$\alpha$-${\cal T}_3$ double layer,  this perturbation
theory is valid only for relatively large separation $D$ between
electron and hole $\alpha$-${\cal T}_3$ double layers
and relatively high magnetic fields $B$, i.e., $D \gg r_{B}$  when
$e^{2}/(\epsilon D) \ll \hbar v_{F}/r_{B}$. Here, $e^{2}/(\epsilon
D)$ is the characteristic Coulomb electron-hole attraction for the
$\alpha$-${\cal T}_3$ double layer  and $\hbar
v_{F}/r_{B}$ is the energy difference  between the first and zeroth
Landau levels in $\alpha$-${\cal T}_3$. The operator
of electron-hole Coulomb attraction is

\begin{eqnarray}\label{el-ho}
\hat{V}(r) = -\frac{k e^{2}}{\epsilon\sqrt{r^{2} + D^{2}}}\  ,
\end{eqnarray}
where  $k=9\times 10^{9}\ N\times m^{2}/C^{2}$, $\epsilon$ is the dielectric constant of the insulator (SiO$_2$ or $h$-BN), surrounding the electron and hole $\alpha$-${\cal T}_3$ monolayers, forming the double layer; $D$ is the separation between electron and hole $\alpha$-${\cal T}_3$ mono-layers.  For the $h$-BN barrier we substitute the dielectric constant $\epsilon= 4.89$, while for the SiO$_2$ barrier we substitute the dielectric constant $\epsilon=4.5$.

\medskip
\par
The magnetoexciton energies $E_{n_{+},n_{-}}(P)$ in first order perturbation theory are  given by

\begin{eqnarray}\label{en1}
E_{n_{e},n_{h}}(P) = E_{n_{e},n_{h}}^{(0)} +
\mathcal{E}_{n_{e},n_{h}}(P)\  ,
\end{eqnarray}
where $E_{n_{e},n_{h}}^{(0)}$ is the unperturbed spectrum, and

\begin{eqnarray}\label{en2}
\mathcal{E}_{n_{e},n_{h}}(P) = - \left\langle
n_{e}n_{h}\mathbf{P}\left| \frac{k e^{2}}{\epsilon\sqrt{D^{2}+
r^{2}}} \right|n_{e}n_{h}\mathbf{P}\right\rangle \ .
\end{eqnarray}
Neglecting the transitions between different Landau levels,  first order perturbation woth respect to the Coulomb attraction leads to the following result for the energy of magnetoexciton  for case 1:

\begin{eqnarray}\label{en3}
 \mathcal{E}_{1,1}(P) = - \left\langle
1,1,\mathbf{P}\left| \frac{k e^{2}}{\epsilon\sqrt{D^{2}+ r^{2}}}
\right|1,1,\mathbf{P}\right\rangle ,
\end{eqnarray}
and for case 2:

\begin{eqnarray}\label{en33}
 \mathcal{E}_{1,0}(P) = - \left\langle
1,0,\mathbf{P}\left| \frac{k e^{2}}{\epsilon\sqrt{D^{2}+ r^{2}}}
\right|1,0,\mathbf{P}\right\rangle \  .
\end{eqnarray}

\medskip
\par

Denoting the averaging involving the 2D harmonic oscillator eigenfunctions $\Phi_{n_{e},n_{h}}(\mathbf{r})$ in Eq.~(\ref{electron_S}) as $\langle\langle \tilde{n}m \mathbf{P}|\cdots| \tilde{n}m \mathbf{P}\rangle\rangle$ ($\tilde{n}$ and $m$ are defined below Eq.~(\ref{electron_S})), we obtain the energy of an indirect magnetoexciton created by spatially separated electrons and holes in the lowest Landau level for case 1:

\begin{eqnarray}\label{en4}
\mathcal{E}_{1,1} (P) &=&  \left\langle
1,1,\mathbf{P}\left|\hat{V}(r)\right|1,1,\mathbf{P}\right\rangle  =
\frac{1}{4} \left[ \left\langle\left\langle
0,0,\mathbf{P}\left|\hat{V}(r)\right|0,0,\mathbf{P}\right\rangle
\right\rangle \right. \nonumber\\
&+& \left. 2 \left\langle\left\langle
0,1,\mathbf{P}\left|\hat{V}(r)\right|0,1,\mathbf{P}\right\rangle
\right\rangle  + \left\langle\left\langle
1,0,\mathbf{P}\left|\hat{V}(r)\right|1,0,\mathbf{P}\right\rangle
\right\rangle \right]
\end{eqnarray}
and for case 2, we have

\begin{eqnarray}\label{en44}
\mathcal{E}_{1,0} (P) &=&  \left\langle
1,0,\mathbf{P}\left|\hat{V}(r)\right|1,0\mathbf{P}\right\rangle  =
\frac{1}{2} \left[ \left\langle\left\langle
0,0,\mathbf{P}\left|\hat{V}(r)\right|0,0,\mathbf{P}\right\rangle
\right\rangle \right. \nonumber \\
 &+& \left. \left\langle\left\langle
0,1,\mathbf{P}\left|\hat{V}(r)\right|0,1,\mathbf{P}\right\rangle
\right\rangle \right] \  .
\end{eqnarray}
Substituting for small magnetic momenta $P \ll \hbar/r_{B}$ and $P \ll \hbar D/r_{B}^{2}$ the following relations  \cite{Ruvinsky}

\begin{eqnarray}\label{ruv}
\langle\langle \tilde{n}m \mathbf{P}|\hat{V}(r)| \tilde{n}m
\mathbf{P}\rangle\rangle =  \mathcal{E}_{\tilde{n}m}^{(b)} +
\frac{P^{2}}{2M_{\tilde{n}m}(B,D)} \  .
\end{eqnarray}
Making use of this in  Eqs.~(\ref{en4}) and~(\ref{en44}), we obtain the dispersion law ofa magnetoexciton for small magnetic momenta in cases 1 and 2, correspondingly, i.e.,

\begin{eqnarray}\label{en5}
\mathcal{E}_{1,1} (P) &=&
\frac{1}{4}\left(\mathcal{E}_{00}^{(b)}(B,D) + 2
\mathcal{E}_{01}^{(b)}(B,D) + \mathcal{E}_{10}^{(b)}(B,D)\right)
\nonumber \\
&+& \frac{1}{4}\left(\frac{1}{M_{00}(B,D)} + \frac{2}{M_{01}(B,D)} + \frac{1}{M_{10}(B,D)}\right) \frac{P^{2}}{2} \nonumber \\
&=& - \mathcal{E}_{B}^{(b)}(D) + \frac{P^{2}}{2m_{B}(D)}\  ,
\end{eqnarray}
and

\begin{eqnarray}\label{en55}
\mathcal{E}_{1,0} (P) &=&
\frac{1}{2}\left(\mathcal{E}_{00}^{(b)}(B,D) +
\mathcal{E}_{01}^{(b)}(B,D)\right) \nonumber \\
&+&  \frac{1}{2} \left(\frac{P^{2}}{2M_{00}(B,D)} +
\frac{P^{2}}{2M_{01}(B,D)}\right) = - \mathcal{E}_{B}^{(b)}(D) +
\frac{P^{2}}{2m_{B}(D)}\  ,
\end{eqnarray}
where the binding energy $\mathcal{E}_{B}^{(b)}(D)$ and the effective magnetic mass $m_{B}(D)$ of a magnetoexciton with spatially separated electron and hole in the $\alpha$-${\cal T}_3$ double layer are for case 1

\begin{eqnarray}\label{m1}
\mathcal{E}_{B}^{(b)}(D) &=& - \frac{1}{4}\left(
\mathcal{E}_{00}^{(b)}(B,D) + 2 \mathcal{E}_{01}^{(b)}(B,D) +
\mathcal{E}_{10}^{(b)}(B,D)\right), \nonumber
\\ \frac{1}{m_{B}(D)} &=& \frac{1}{4}\left(\frac{1}{M_{00}(B,D)} +
\frac{2}{M_{01}(B,D)} + \frac{1}{M_{10}(B,D)}\right),
\end{eqnarray}
and for case 2:

\begin{eqnarray}\label{m11}
\mathcal{E}_{B}^{(b)}(D) &=& - \frac{1}{2} \left(\mathcal{E}_{00}^{(b)}(B,D) +  \mathcal{E}_{01}^{(b)}(B,D) \right) , \nonumber \\
\frac{1}{m_{B}(D)} &=& \frac{1}{2} \left(\frac{1}{M_{00}(B,D)} +
\frac{1}{M_{01}(B,D)}\right),
\end{eqnarray}
 where the constants
$\mathcal{E}_{00}^{(b)}(B,D)$, $\mathcal{E}_{01}^{(b)}(B,D)$,
$\mathcal{E}_{10}^{(b)}(B,D)$, $M_{00}(B,D)$, $M_{01}(B,D)$ and
$M_{10}(B,D)$ depending on magnetic field $B$ and the inter-layer
separation $D$ are defined by  \cite{Ruvinsky}

\begin{eqnarray}\label{m2}
\mathcal{E}_{00}^{(b)}(B,D) &=&
-\mathcal{E}_{0}\exp\left[\frac{D^{2}}{2r_{B}^{2}}\right]
 \mathrm{erfc}\left[\frac{D}{\sqrt{2}r_{B}}\right] , \nonumber \\
 \mathcal{E}_{01}^{(b)}(B,D) &=& -\mathcal{E}_{0}\left[\left(\frac{1}{2} - \frac{D^{2}}{2r_{B}^{2}} \right)\exp\left[\frac{D^{2}}{2r_{B}^{2}}\right] \mathrm{erfc}\left[\frac{D}{\sqrt{2}r_{B}}\right] + \frac{D}{\sqrt{2\pi}r_{B}}
  \right], \nonumber \\
  \mathcal{E}_{10}^{(b)}(B,D) &=& -\mathcal{E}_{0}\left[\left(\frac{3}{4} + \frac{D^{2}}{2r_{B}^{2}} + \frac{D^{4}}{4r_{B}^{4}} \right)\exp\left[\frac{D^{2}}{2r_{B}^{2}}\right] \mathrm{erfc}\left[\frac{D}{\sqrt{2}r_{B}}\right] - \frac{D}{2\sqrt{2\pi}r_{B}}- \left(\frac{D}{\sqrt{2}r_{B}}\right)^{3}\frac{1}{\sqrt{\pi}}
  \right], \nonumber \\
 M_{00}(B,D) &=&  M_{0}\left[\left(1+\frac{D^{2}}{r_{B}^{2}}\right)\exp\left[\frac{D^{2}}{2r_{B}^{2}}\right]
 \mathrm{erfc}\left[\frac{D}{\sqrt{2}r_{B}}\right]
 -\sqrt{\frac{2}{\pi}}\frac{D}{r_{B}}\right]^{-1} , \nonumber \\
 M_{01}(B,D) &=&  M_{0}\left[\left(3+\frac{D^{2}}{r_{B}^{2}}\right)\frac{D}{\sqrt{2\pi} r_{B}}
 -  \left(\frac{1}{2} + 2\frac{D^{2}}{r_{B}^{2}} + \frac{D^{4}}{2r_{B}^{4}}\right)\exp\left[\frac{D^{2}}{2r_{B}^{2}}\right]
 \mathrm{erfc}\left[\frac{D}{\sqrt{2}r_{B}}\right]\right]^{-1} , \nonumber \\
 M_{10}(B,D) &=& M_{0}\left[\frac{1}{4}\left(7 + 25\frac{D^{2}}{r_{B}^{2}} + 11 \frac{D^{4}}{r_{B}^{4}}+ \frac{D^{6}}{r_{B}^{6}}\right)\exp\left[\frac{D^{2}}{2r_{B}^{2}}\right]\mathrm{erfc}\left[\frac{D}{\sqrt{2}r_{B}}\right] \right. \nonumber \\&-& \left.  \left(\frac{17}{2} + 5\frac{D^{2}}{r_{B}^{2}} + \frac{D^{4}}{2r_{B}^{4}}\right)\frac{D}{\sqrt{2\pi} r_{B}}\right]^{-1} \ ,
\end{eqnarray}
where the constants $\mathcal{E}_{0}$ and $M_{0}$ and function $\mathrm{erfc}(z)$ are given by  \cite{Ruvinsky}

\begin{eqnarray}\label{m0}
\mathcal{E}_{0} &=& \left\langle\left\langle 00 \mathbf{P}\left|\frac{e^{2}}{\epsilon|\mathbf{r}|}\right| 00 \mathbf{P}\right\rangle\right\rangle_{\mathbf{P}=0} = \frac{k e^{2}}{\epsilon r_{B}}\sqrt{\frac{\pi}{2}}, \nonumber \\
M_{0} &=& -\left[2\left(\left\langle\left\langle 00 \mathbf{P}\left|\frac{e^{2}}{\epsilon|\mathbf{r}|}\right| 00 \mathbf{P}\right\rangle\right\rangle - \mathcal{E}_{0}\right)\right]^{-1}P^{2} = \frac{2^{3/2}\epsilon \hbar^{2}}{\sqrt{\pi}k e^{2}r_{B}}, \nonumber \\
\mathrm{erfc}(z)&=&
\frac{2}{\sqrt{\pi}}\int_{z}^{\infty}\exp\left(-t^{2}\right)dt \ .
\end{eqnarray}

For  both cases 1 and 2, for large inter-layer separation $D\gg r_B$, the asymptotic values for the binding energy $\mathcal{E}_{B}^{(b)}(D)$ and the effective magnetic mass $m_B(D)$  of the dipolar magnetoexciton in  the $\alpha$-${\cal T}_3$ double layer   are the same and given by

\begin{eqnarray}\label{as1}
\mathcal{E}_{B}^{(b)}(B,D) =   \frac{k e^2}{\epsilon D}, \ \ \
m_B(D) = \frac{\epsilon D^{3}B^2}{k} \  .
\end{eqnarray}

\medskip
\par

Measuring energy from the binding energy of the  magnetoexciton, the dispersion relation  $\varepsilon _{k}(P)$  for an isolated dipolar magnetoexciton is a quadratic function at small magnetic momentum $P \ll \hbar/r_{B}$ and  $P \ll \hbar D/r_{B}^{2}$:

 \begin{equation}\label{Energy}
\varepsilon _{k}({\bf P}) = \frac{P^2}{2m_{B k}}\ ,
\end{equation}
where $m_{B k}$, the effective magnetic mass, dependent  on $B$ and the separation $D$ between electron and hole layers as well as the quantum number $k$ ($k = (n_{e},n_{h})$ are magnetoexcitonic quantum numbers).

\medskip
\par

The squared 2D radius of a magnetoexciton for case 1 can be defined as

\begin{eqnarray}\label{ln}
r_{1,1}^{2} (P=0) &=& \left\langle 1,1,\mathbf{P}\left|r^{2}
\right|1,1,\mathbf{P}\right\rangle_{\mathbf{P} = 0}  =
\frac{1}{4}\left(l_{00}^{2}+2l_{01}^{2}+l_{10}^{2}\right) = 4
r_{B}^{2}, \hspace{1cm} r_{1,1} = 2r_{B} ,
\end{eqnarray}
and for case 2 as

\begin{eqnarray}\label{ln1}
r_{1,0}^{2} (P=0) &=& \left\langle 1,0,\mathbf{P}\left|r^{2}
\right|1,0,\mathbf{P}\right\rangle_{\mathbf{P} = 0}  =
\frac{1}{2}\left(l_{00}^{2}+ l_{01}^{2}\right) = 3 r_{B}^{2},
\hspace{1cm} r_{1,0} = \sqrt{3}r_{B} \ ,
\end{eqnarray}
 where $l_{\tilde{n}m}^{2} = \left\langle\left\langle \tilde{n}m
\mathbf{P}\left|\hat{r^{2}}\right| \tilde{n}m
\mathbf{P}\right\rangle\right\rangle_{\mathbf{P}=0} $ and
($l_{00}^{2}=2r_{B}^{2}$, $l_{01}^{2}=4r_{B}^{2}$,
$l_{10}^{2}=6r_{B}^{2}$)~\cite{Ruvinsky}.

\section{Superfluidity of dipolar magnetoexcitons in an $\alpha$-${\cal T}_3$  double layer}
\label{sec5}

 Dipolar magnetoexcitons  have electrical dipole moments, produced by the inter-layer separation  $D$.
 We assume, that dipolar magnetoexcitons repel each other like parallel dipoles.
 The latter assumption is reasonable  when $D$ is larger than the mean separation between an electron and hole
 parallel to the $\alpha$-${\cal T}_3$ layers  $D \gg \left(\langle r^{2} \rangle\right)^{1/2}$.

\medskip
\par

Since electrons on an $\alpha$-${\cal T}_3$ monolayer can be located in
 two valleys, there are four types of dipolar magnetoexcitons in an $\alpha$-${\cal T}_3$ double layer.
 Since all these types of dipolar magnetoexcitons have identical envelope wave functions and energies,
 it is reasonable to assume that a dipolar magnetoexciton is located in only one valley. We use $n_{0} =n/(4s)$
 as the density of magnetoexcitons in one valley, where $n$ is the total density of magnetoexcitons, $s$ is the spin degeneracy ( $s=4$ for magnetoexcitons in an $\alpha$-${\cal T}_3$ double layer).

\medskip
\par

We shall treat a dilute  2D  magnetoexciton system in the $\alpha$-${\cal T}_3$ double layer as a weakly interacting Bose gas by applying the procedure, described in Ref.~\cite{BLG}. Two dipolar magnetoexcitons in a dilute system repel each other with the potential energy of the pair magnetoexciton-magnetoexciton interaction, written as $U(R) = k e^{2}D^{2}/(\epsilon R^{3})$, where $R$ is the distance between magnetoexciton dipoles along the $\alpha$-${\cal T}_3$ layers. For the weakly interacting Bose gas of 2D dipolar magnetoexcitons  (when $n a^{2}(B) \ll 1$, where $a(B)$ is the in-plane radius of a dipolar magnetoexciton defined for cases 1 and 2  in Eqs.~(\ref{ln}) and~(\ref{ln1}), respectively)  the summation of ladder diagrams is valid~\cite{Abrikosov}. The chemical potential $\mu $, corresponding to the summation of the ladder diagrams, can be written as~\cite{BLG}

\begin{eqnarray}
\label{Mu}
\mu =  \frac{\kappa ^2 }{2m_{B}} = \frac{\pi \hbar^{2}n}{sm_B \log
\left[ s\hbar^{4}\epsilon^{2}/\left(2\pi n m_B^2 k^{2} e^4
D^4\right) \right]} \  ,
\end{eqnarray}
where $s=4$ is the spin degeneracy factor.

\medskip
\par

 The spectrum of collective excitations, obtained from the ladder approximation, at low magnetic momenta corresponds  to the sound spectrum of collective excitations $\varepsilon (P)  = c_s P$ with  the sound velocity $c_s =  \sqrt{\mu/m_B}$, where $\mu $ is defined by Eq.~(\ref{Mu}). Since magnetoexcitons have a sound spectrum for collective excitations at small magnetic momenta $P$ due to dipole-dipole repulsion, the magnetoexcitonic superfluidity is possible at low temperatures $T$ in $\alpha$-${\cal T}_3$ double layers because the sound spectrum satisfies the Landau criterion for superfluidity \cite{Abrikosov,Griffin}.

\medskip
\par

The magnetoexcitons constructed from spatially separated electrons and holes in the $\alpha$-${\cal T}_3$ double layers with large inter-layer separation $D \gg r_{B}$ form a weakly interacting 2D  gas of bosons with a dipole-dipole pair repulsion. Consequently, the  superfluid-normal phase transition in this system is the Kosterlitz-Thouless transition \cite{Kosterlitz}. The temperature $T_c $ of this phase change to the superfluid state in a  2D magnetoexciton system is determined by the equation

\begin{eqnarray}\label{T_KT}
T_c = \frac{\pi \hbar ^2 n_s (T_c)}{2 k_B m_{B}}\  ,
\end{eqnarray}
where $n_s (T)$ is the superfluid density of the magnetoexciton system  as a function of  temperature $T$, magnetic field $B$,  inter-layer separation $D$; and $k_B$ is the Boltzmann constant. The function $n_s (T)$ in Eq.\  (\ref{T_KT}) can be determined from the relation $n_s = n/(4s) - n_n $, where $n$ is the total density, $n_{n}$ is the normal component density. Following the procedure, described in Ref.~[\onlinecite{BLG}], we have for the superfluid density

\begin{eqnarray} \label{n_s} n_s = \frac{n}{4s} - n_n = \frac{n}{4s} -
 \frac{3 \zeta (3) }{2 \pi \hbar^{2}} \frac{k_{B}^{3}T^3}{c_s^4 m_B}\ .
\end{eqnarray}

\medskip
\par
In a 2D system, superfluidity of magnetoexcitons appears below the Kosterlitz-Thouless transition temperature (Eq.~(\ref{T_KT})), where only coupled vortices are present  \cite{Kosterlitz}. Using Eq.~(\ref{n_s}) for the density $n_{s}$ of the superfluid component, we obtain an equation for the Kosterlitz-Thouless transition temperature $T_{c}$. Its solution is

\begin{eqnarray}
\label{tct} T_c = \left[\left( 1 +
\sqrt{\frac{32}{27}\left(\frac{sm_{B}k_{B}T_{c}^{0}}{\pi \hbar^{2} n}\right)^{3} +
1} \right)^{1/3}   - \left( \sqrt{\frac{32}{27} \left(\frac{sm_{B}k_{B}T_{c}^{0}}{\pi \hbar^{2} n}\right)^{3} + 1} - 1 \right)^{1/3}\right]
\frac{T_{c}^{0}}{ 2^{1/3}} \  .
\end{eqnarray}
Here,  $T_{c}^{0}$ is the temperature at which the superfluid density vanishes in the mean-field approximation (i.e., $n_{s}(T_{c}^{0}) = 0$),

\begin{equation}
\label{tct0} T_c^0 = \frac{1}{k_{B}} \left( \frac{ \pi \hbar^{2} n c_s^4 m_{B}}{6 s\zeta (3)}
\right)^{1/3} \ .
\end{equation}

\begin{figure}[h!]
\centering
\includegraphics[width=0.85\textwidth]{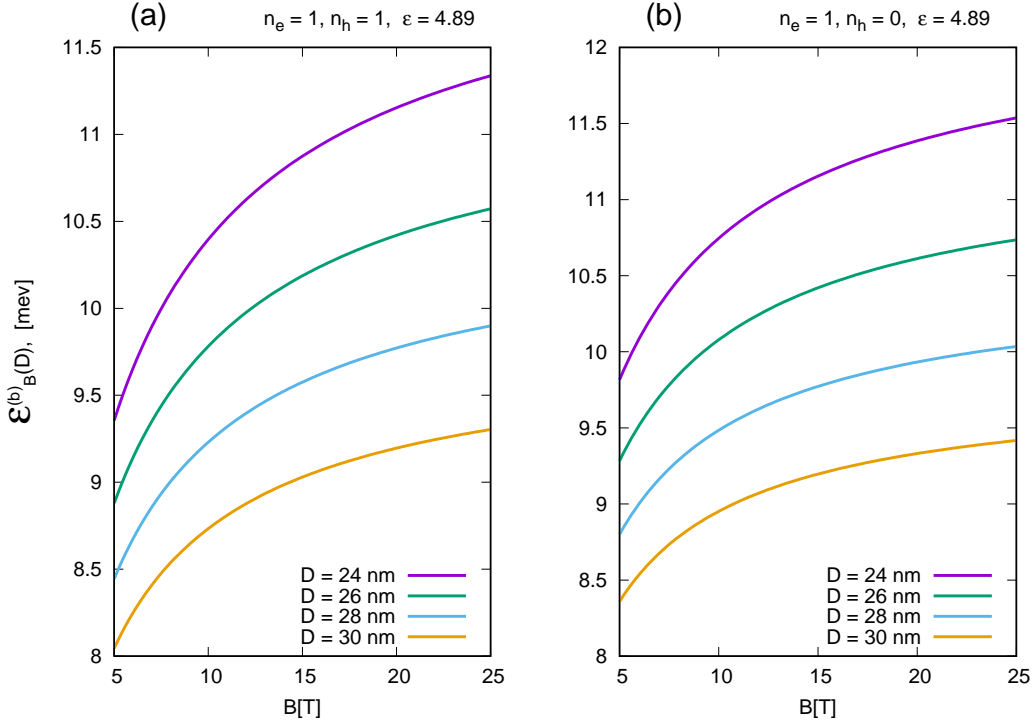}
\caption{(Color online) \  The magnetoexciton binding energy $\mathcal{E}_{B}^{b}(B,D)$ as a function of magnetic field $B$ for chosen interlayer separations $D$ for (a) case 1 on the left-hand side and (b) case 2, on the right.}
\label{Fig:2}
\end{figure}

\medskip
\par
In Fig.\  \ref{Fig:2}, we present results showing the dependence of the magnetoexciton binding energy $\mathcal{E}_{B}^{b}(B,D)$ on the magnetic field $B$ for chosen interlayer separation $D$ in case 1 and case 2, respectively. According to  Fig.\ \ref{Fig:2},  $\mathcal{E}_{B}^{b}(B,D)$ is increased as $B$ is increased and $D$ is decreased. For the same parameters $\mathcal{E}_{B}^{b}(B,D$ is
slightly larger for case 2 than case 1.

\medskip
\par

\begin{figure}[h!]
\centering
\includegraphics[width=0.85\textwidth]{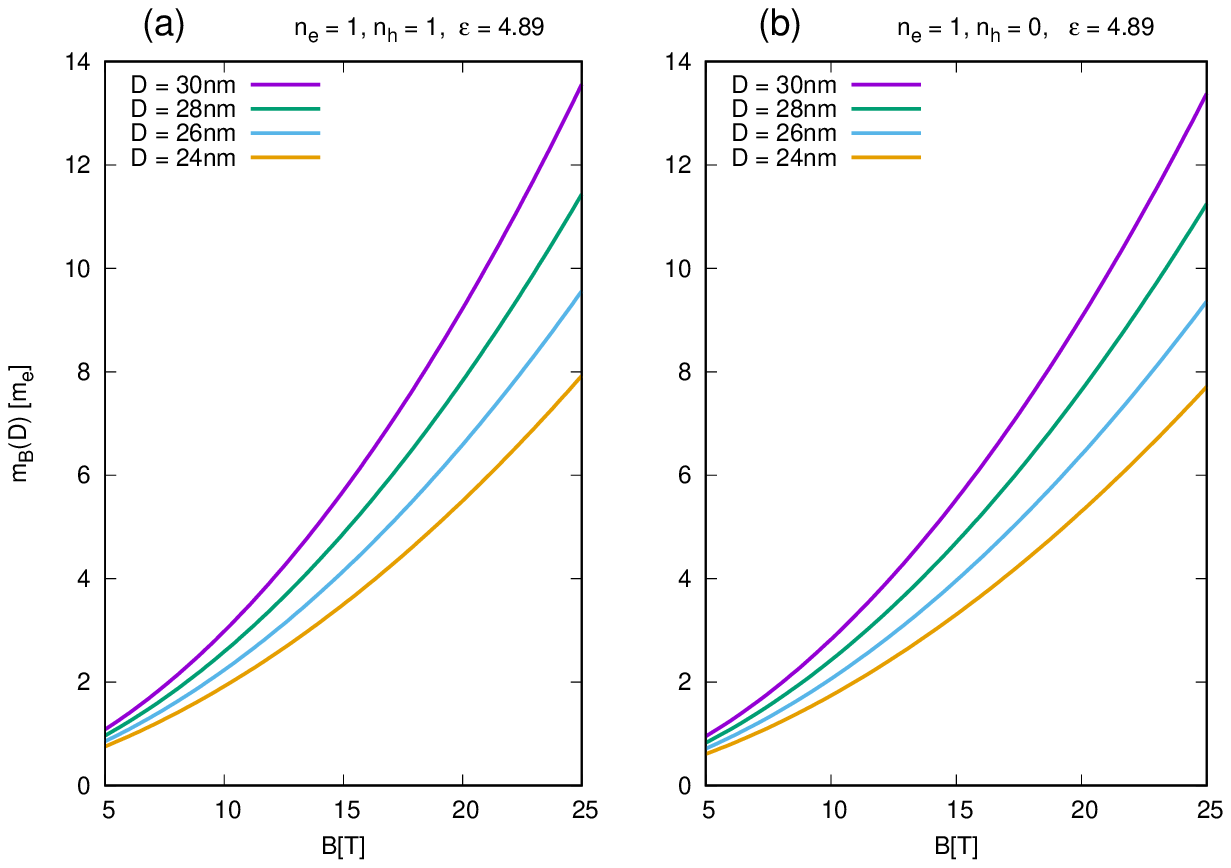}
\caption{(Color online)\  The effective magnetic mass $m_{B}(B,D)$ of a magnetoexciton as a function of magnetic field $B$ for chosen interlayer separation $D$ for (a) case 1 on the left-hand side and (b) case 2, on the right.}
\label{Fig:3}
\end{figure}

Figure  \ref{Fig:3} presents our results for the dependence of the effective magnetic mass $m_{B}(B,D)$ of a magnetoexciton on the magnetic field $B$ for chosen interlayer separation $D$ for cases 1 and 2. According to  Fig.\ \ref{Fig:3}, $m_{B}(B,D)$ is increased as $B$ is increased and $D$ is increased. For the same parameters, $m_{B}(B,D)$ is slightly larger for case 1 compared with case 2.

\begin{figure}[h!]
\centering
\includegraphics[width=0.85\textwidth]{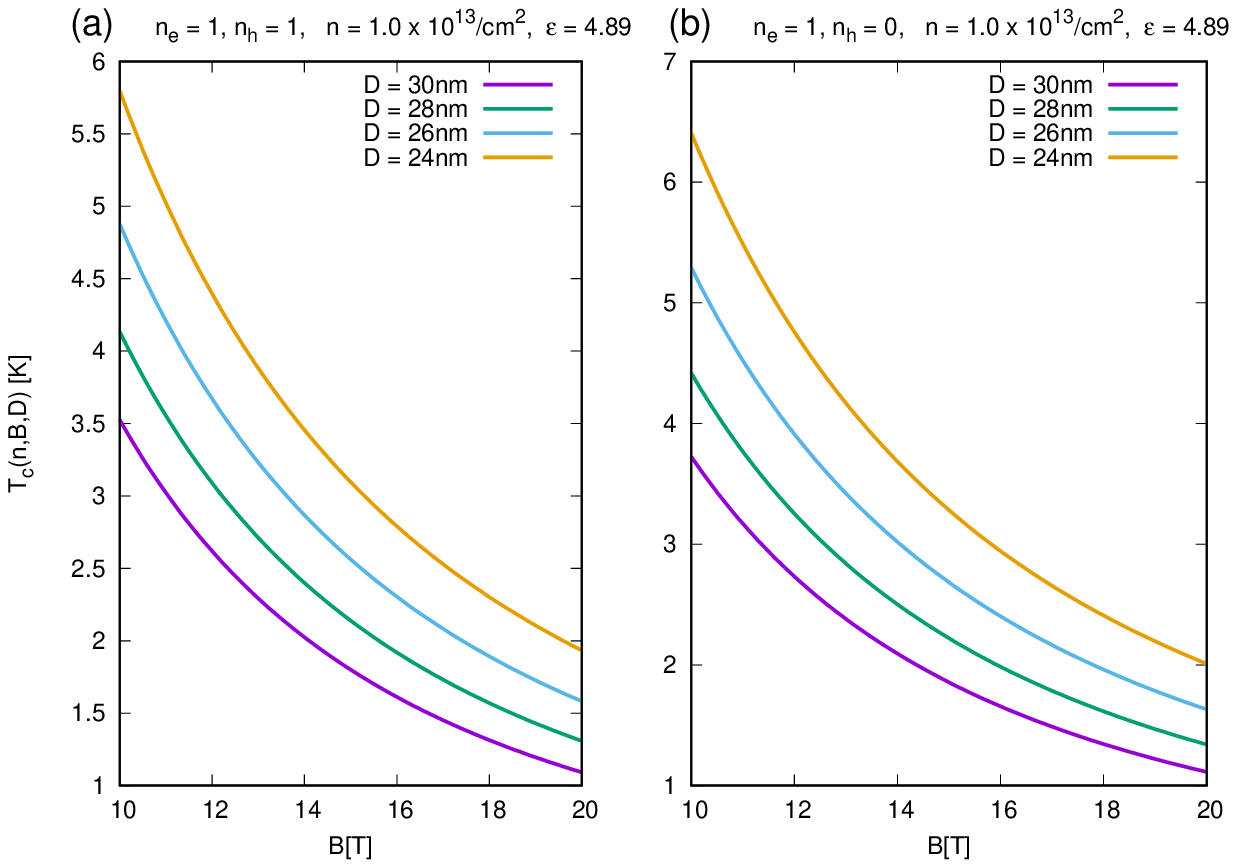}
\caption{(Color online) The Kosterlitz-Thouless transition temperature $T_{c}(n, B,D)$  versus  magnetic field $B$ for chosen interlayer separations $D$ at the fixed magnetoexciton concentration $n$ for (a)  case 1 on the left-hand side and (b) case 2, on the right.}
\label{Fig:4}
\end{figure}

\medskip
\par
In Fig.\ \ref{Fig:4}, we display our results for  the Kosterlitz-Thouless transition temperature $T_{c}(n, B,D)$ versus the magnetic field $B$ for various interlayer separations at fixed magnetoexciton concentration $n$ for cases 1 and 2. According to  Fig.\ \ref{Fig:4}, $T_{c}(n, B,D)$ is decreased as $B$ is increased and $D$ is increased. For the same parameters,  $T_{c}(n, B,D)$ is slightly larger for case 2 compared with case 1.

\begin{figure}[h!]
\centering
\includegraphics[width=0.85\textwidth]{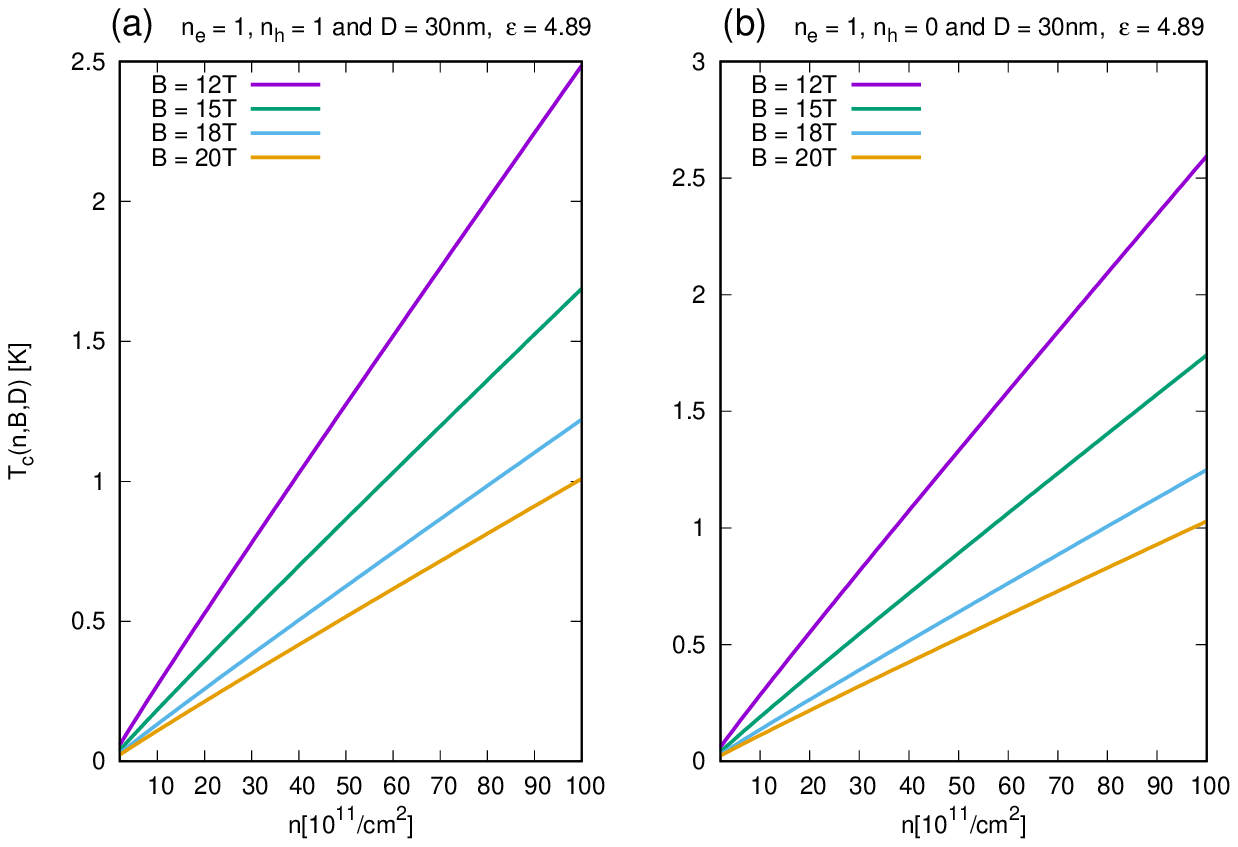}
\caption{(Color online)\  The Kosterlitz-Thouless transition temperature $T_{c}(n, B,D)$ as a function of the magnetoexciton concentration $n$ for various magnetic fields $B$ and fixed interlayer separation $D$ for (a) case 1 on the left-hand side and (b) case 2, on the right.} \label{Fig:5}
\end{figure}

\medskip
\par

We have plotted in Fig.\  \ref{Fig:5}  the functional dependence of the Kosterlitz-Thouless transition temperature $T_{c}(n, B,D)$ on the magnetoexciton concentration $n$ for several chosen magnetic fields $B$ and fixed interlayer separation $D$ in both case 1 and 2.   We deduce from Fig.\ \ref{Fig:5} that $T_{c}(n, B,D)$ is increased as $n$is  increased but is decreases ad as $B$ is increased.  Additionally, we conclude that for the same values of the parameters,  $T_{c}(n, B,D)$ is slightly larger for case 2 than case 1.

\medskip
\par

Based on Figs.~\ref{Fig:2},~\ref{Fig:4}  and~\ref{Fig:5}, one can conclude that case 2 is slightly more preferable than case 1 for observing dipolar magnetoexcitons and their superfluidity  in the $\alpha$-${\cal T}_3$ double layer, since case 2 corresponds to slightly larger magnetoexciton binding energy $\mathcal{E}_{B}^{b}(B,D)$ as well as the Kosterlitz-Thouless transition temperature $T_{c}(n, B,D)$ than case 1 for the same described parameters.

\section{Conclusions}
\label{sec6}

In this paper, we have proposed the occurrence of BEC and
superfluidity of dipolar magnetoexcitons in
$\alpha$-${\cal T}_3$ double layers in a strong
uniform perpendicular  magnetic field. The low-energy
Hamiltonian for a single $\alpha$-${\cal T}_3$ layer was obtained by
including additional hopping terms to a single layer graphene Dirac
Hamiltonian. We have found the solution of a two-body problem for an
electron and a hole for the model Hamiltonian for the
$\alpha$-${\cal T}_3$ double layer in a magnetic field. We have
calculated the binding energy, effective mass, spectrum of
collective excitations, superfluid density and the temperature of
the Kosterlitz-Thouless phase transition to the superfluid state for
dipolar magnetoexcitons in the $\alpha$-${\cal T}_3$
double layer. We have demonstrated that at fixed exciton density,
the Kosterlitz-Thouless temperature for superfluidity of dipolar
magnetoexcitons is decreased as a function of magnetic field. Our
results show that $T_{c}$ is increased as a function of the density
$n$ and is decreased as a function of  the magnetic field $B$ and
the interlayer separation $D$. We have demonstrated that case 2 (the
dipolar magnetoexciton is formed by an electron in  Landau level $1$
and hole in Landau level $0$)
 is slightly more preferable than case 1 (the dipolar magnetoexciton is formed by an electron in Landau level $1$ and hole in Landau level $-1$)
  to observe the dipolar magnetoexcitons and their superfluidity  in $\alpha$-${\cal T}_3$ double layers. The reason is that  case 2
  corresponds to slightly larger magnetoexciton binding energy  and Kosterlitz-Thouless transition temperature than case 1 for the same chosen parameters.

\end{document}